\documentstyle[11pt,fleqn]{article}
\begin{document}
\newcommand{\newc}{\newcommand}

\newc{\be}{\begin{equation}}
\newc{\ee}{\end{equation}}
\newc{\ba}{\begin{eqnarray}}
\newc{\ea}{\end{eqnarray}}
\newc{\ie}{{\it i.e.}}
\newc{\eg}{{\it eg.}}
\newc{\etc}{{\it etc.}}
\newc{\etal}{{\it et al.}}

\newc{\ra}{\rightarrow}
\newc{\lra}{\leftrightarrow}
\newc{\no}{Nielsen-Olesen }
\newc{\lsim}{\buildrel{<}\over{\sim}}
\newc{\gsim}{\buildrel{>}\over{\sim}}

\begin{titlepage}
November 1993\hfill
BROWN-HET-926
\begin{center}

\vskip 1in

{\large \bf
Signatures of Cosmic Strings on the Microwave Background
}
\vskip .4in
{\large L. Perivolaropoulos\footnote{Division of Theoretical Astrophysics,
Harvard-Smithsonian Center for Astrophysics,
60 Garden St.,
Cambridge, Mass. 02138.}$^{,2}$\hspace{.2cm} R. Moessner\footnote{
Department of Physics, Brown University, Providence, R.I. 02912.}
 \hspace{.2cm} R. Brandenberger$^2$}
\vskip 1cm
{\large \it To appear in the proceedings of the Capri MBR Workshop \\
Capri (Italy), September 20-24 1993}
\end{center}

\vskip .2in
\begin{abstract}
\noindent
We review recent progress on testing the hypothesis of the existence of cosmic
string
perturbations in microwave background maps.
Using an analytical model for the string network we show that the predicted
amplitude and spectrum of
MBR fluctuations are consistent with the COBE data for a reasonable value of
the
single free parameter
of the string model (the mass per  unit length of the string $\mu$). To {\it
distinguish} the predictions of cosmic strings from those of Gaussian models it
is necessary to apply specific {\it
statistical tests} which are sensitive to non-random phases of the MBR
temperature maps. We discuss two such
statistical tests: First, the probability distribution and the moments of
fluctuations ${{\delta
T}\over T}$ and second the corresponding statistical quantities for the {\it
gradient} of fluctuations
along a fixed direction. We show that the second statistic can detect cosmic
string specific signatures on MBR temperature maps with resolution of a few
arcminutes or smaller.

\end{abstract}

\end{titlepage}

\section{Introduction}
The quest for an understanding of the origin of Microwave
Background Radiation (MBR) anisotropies has become a central open issue since
their recent discovery.
There are three main candidate types of primordial fluctuations that may have
caused these
anisotropies. First, Gaussian perturbations produced by quantum fluctuations of
a scalar field during
inflation. Second, seed perturbations which are naturally generated by
topological defects produced
during a phase transition in the early universe. Third, other types of
non-Gaussian perturbations
produced for example by special inflationary potentials or late time phase
transitions. Here we
review recent progress
on testing for seed-like perturbations. In particular we focus on the case
of cosmic strings (Kibble 1976; Vilenkin 1985; Brandenberger 1992) and address
the following
questions:
\begin{itemize}
\item
Are the cosmic string predictions consistent with the COBE data?
\item
What statistical tests can identify cosmic string signatures on MBR temperature
maps?
\end{itemize}
In order to address these questions we use a simple analytic model (Vachaspati
1992) for the effects of a distribution of cosmic strings
which was first applied to the study of MBR fluctuations by
Perivolaropoulos (1993a) (see also Hara \& Miyoshi 1993) . Some of the main
results
obtained using this model include the predicted amplitude and spectrum of MBR
fluctuations on COBE angular scales and the predicted probability distribution
and moments of the
temperature fluctuations ${{\delta T} \over T}$ and their {\it gradient}. It
turns out that the
statistical study of the {\it gradient} of  temperature fluctuations offers a
powerful probe of the
signatures of cosmic strings on scales of a few arcminutes or smaller (Moessner
et. al. 1994). Other
interesting statistical tests  of non-Gaussianity have also been proposed
recently (see e.g. Luo 1993;
Graham et. al. 1993; Coulson et. al. 1993; Perivolaropoulos 1993c).

The main assumptions on which the analytical model is based are the following:
First, long strings,
present after the time of recombination are assumed to dominate perturbations
(this is also
suggested in the string evolution simulations of Bouchet et. al. (1988)).
Second, strings are
approximately straight on horizon scales (Bennett \& Bouchet 1990, Allen \&
Shellard 1990). Third,
string positions and velocities are uncorrelated on  Hubble space- and time-
scales (by causality).

The method used is based on a discretization of the photon path from the time
of
recombination to today into a
set of ``Hubble" time-steps (Fig. 1). The first step begins at recombination
when the horizon
scale is of the order $t_{rec}$ and each subsequent time step occurs when the
horizon has doubled in
size. Each individual string present in a given Hubble slice produces a
Kaiser-Stebbins
temperature kick (Kaiser \& Stebbins 1984; Stebbins 1988) equal to $\pm 4\pi
G\mu\beta$ which is effective
only within a horizon distance from the string (Traschen et.al. 1986;
Veeraraghavan \& Stebbins 1990, Magueijo 1992). Here $\beta={\hat
k}\cdot({\vec v_s}\gamma_s \times {\hat s})$ where $v_s$ is the magnitude of
the
average string
velocity, $\hat s$ is a unit vector along the string and $\hat k$ is the unit
photon wave-vector.  It
is then  straightforward to add up the contributions of all strings
(Perivolaropoulos 1993a) using
the assumption that there is no
correlation among strings in  different Hubble time-steps and Hubble volumes.

The model offers a fairly powerful analytical tool to compute several
observable
predictions of seed-based models. For example it is straightforward to obtain
analytical expressions for the
temperature correlation function $C(\alpha)$ (Perivolaropoulos 1993a), the
probability distribution
of temperature  fluctuations (Perivolaropoulos 1993b) or their gradient
(Moessner et. al. 1994),
properties of peculiar velocities (Vachaspati 1992; Perivolaropoulos \&
Vachaspati 1993) {\it etc}.

\section{Amplitude, Spectrum}
Numerical simulations (Bennett \& Bouchet 1990, Allen \& Shellard 1990) have
shown clear evidence
that the evolving cosmic string network rapidly approaches a scale invariant
configuration called the
{\it scaling solution}. According to the scaling solution there are a fixed
number of long strings per
horizon volume per Hubble time moving with relativistic velocities. There is
also a distribution of loops smaller than the Hubble radius whose statistical
properties do not depend on time if all lengths are scaled to the horizon.
This
scale invariant string configuration produces density fluctuations and induced
MBR anisotropies which are also scale invariant
and may therefore be approximately described by a Harrisson-Zeldovich (n = 1)
spectrum. Thus, to a first approximation the shape of the spectrum is the same
as that predicted in the simplest inflationary Universe models.

A more quantitative
study of predicted temperature fluctuations may be obtained using the
analytic model described in the previous section.
In this way, it is straightforward to obtain an expression for the angular
autocorrelation function $C(\alpha)$ of temperature fluctuations,
valid on angular scales larger than about a degree. The result is
(Perivolaropoulos 1993a)
$$
C(\alpha)={\xi^2\over 3} \cos \alpha (\log_2 ({t_0\over t_{rec}})-3 \log_2 (1+
{\alpha \over \alpha_{rec}}))
\eqno (1)
$$
where $\xi\equiv 4\pi G\mu v_s \gamma_s \sqrt{M}$, $v_s$ is the magnitude of
the
string velocity
and $M$ is the number of strings per Hubble volume. Fig. 2 shows a plot of
$C(\alpha)$
smoothed on angular scales of $10^\circ$ (as in Smoot {\it et. al.} (1992))
with
$v_s \gamma_s = 0.15$
and $M=10$ as suggested by string network numerical simulations (Allen \&
Shellard (1990)).

 The units for the curve that corresponds to strings are
$(\mu_6\hskip 1mm (\mu K))^2$ where $\mu_6\equiv {{G\mu} \over 10^{-6}}$ is the
single free parameter
of the cosmic string model. A value of $\mu_6\simeq 1$ is required
in order to obtain reasonable agreement of cosmic string predictions with large
scale structure
(Stebbins et. al. 1987; Perivolaropoulos et. al. 1990), galaxy observations
(Turok \& Brandenberger
1986) and the temperature autocorrelation function obtained by the DMR of COBE
(Smoot {\it et. al.} (1992)).
In addition, such a value of $\mu_6$ is consistent with constraints
coming from particle physics
if cosmic strings were produced during a grand unification phase transition.

The best fit to the COBE data can be obtained for $\mu_6 \simeq 1.7 \pm 0.7$
(Perivolaropoulos 1993a) while the effective power spectrum index
($n =1.1\pm 0.4$) is consistent
with a scale
invariant Zeldovich spectrum. As mentioned above, such a spectrum could have
been expected since the
perturbations were constructed in a scale invariant way by the string scaling
solution.
These results are in good agreement with other recent studies based on
numerical
simulations of string
evolution (Bennett et. al. (1992)).

\section{Statistical Properties of Fluctuations}
The predictions for the correlation function can test the cosmic string model
by
comparing its
predictions of amplitude and spectrum of perturbations with the COBE data but
can not
distinguish it from other theories which are also consistent with COBE. This
distinction can be made by identifying the non-Gaussian properties of string
induced perturbations.
The simplest test of non-Gaussianity is provided by calculating the probability
distribution and the
moments of ${{\delta T}\over T}$.

It is straightforward to use the analytical model described above to find
expressions for the
probability distributions and the moments of string induced perturbations
(Perivolaropoulos 1993b).
The large number of strings per horizon volume implies (by the central limit
theorem)
that the  resulting perturbations are approximately Gaussian. This is shown in
Fig. 3a where we plot
the  probability distribution $P({{\delta T}\over T})$ for strings vs a
variable
proportional to
${{\delta T}\over T}$. The maximum relative difference from the corresponding
Gaussian probability
distribution is about 1\% even though the existence of long non-Gaussian tails
is clear (Fig. 3b).
The predicted value of the skewness is 0 due to the symmetry between positive
and
negative perturbations in the mechanism that creates them. The relative
difference of the kurtosis
from the Gaussian varies inversely proportional to the number of strings $M$
per
Hubble volume
and is less than 1\%.
These results show that the predicted difference from the Gaussian for this
statistic is negligible.
 It may be shown by using the central limit theorem
that the effects of smoothing (not taken into account in the above analysis)
always tend to increase
further the Gaussian character of fluctuations.

For comparison, in the case of textures this test is more useful due to the
small number of knots
unwinding per Hubble volume. There are only 0.04 textures collapsing per Hubble
volume per Hubble
time while the corresponding number for strings is 10.  It may be shown
(Perivolaropoulos 1993b) that
the maximum relative difference of the probability distribution from the
Gaussian for textures
is  about 10\% (without smoothing) while the relative deviation of the kurtosis
from the Gaussian is
about 30\%. These deviations however become negligible when smoothing on COBE
scales
is taken into account.

Fortunately, there is a much more powerful test that can identify the
non-Gaussian features
due to strings. It involves the study of the statistics of temperature
differences (gradients)
rather than temperature fluctuations. The model described above can be applied
to obtain the
temperature differences between neighboring pixels in an MBR experiment. Since
each long string
produces a temperature discontinuity, the pattern of temperature differences
will consist of
a superposition of localized spikes. Such a pattern approaches much more slowly
the Gaussian
for a large number of spikes than a superposition of step-functions which would
be the temperature
fluctuation pattern. Fig. 4 shows the predicted probability distribution for an
experiment with
resolution 18'' (e.g the VLA experiment (Fomalont et. al. 1993)). Superimposed
are the corresponding
Gaussian distribution with the same variance and the result of a Monte-Carlo
simulation confirming
the analytical result. The value of the kurtosis predicted for such an
experiment is a
factor of approximately 3 larger than the Gaussian value $k_g=3$.

Since each experiment only makes a finite  number of measurements, it can not
measure the true
value of moments. The probability distribution of the measured moments about
their true
values will be a Gaussian due to the central limit theorem. The variance
$\sigma_{k_4}^2$ of the
kurtosis $k_4$ depends on the number of measurements $n$ as
$\sigma_{k_4}^2={1\over
n}(k_8-k_4^2)$ where $k_8$ and $k_4$ are the normalized 8th and 4th moments
respectively
(Moessner et. al. (1994)).
 Fig. 5 shows the predicted probability distribution of the kurtosis
for the VLA experiment in the cases of underlying
Gaussian
and stringy  perturbations. For a beam size of 18'', due to the large number of
pixels ($n\simeq 1000$),
the overlap of the Gaussian and stringy distributions is small enough to allow
a
clear distinction
between the models. This is not so for a beam size of 80''. In this case,
even though the
predicted kurtosis is about 50\% larger than the Gaussian, the overlap of the
two distributions is too
large to allow definite conclusions.

In conclusion, there are two main points supported by the work reviewed here:
First, the cosmic string
model is consistent with the COBE data and second there are well defined
statistical tests
that can probe non-Gaussian features induced by strings on scales of arcminutes
or smaller.

\section{Acknowledgements}
\par
This work was supported by a CfA Postdoctoral Fellowship (L.P.) and by DOE
grant
FG02-91ER40688, Task A.

\section{Figure Captions}
\begin{flushleft}
{\bf Figure 1 :} The discretization of the photon path from the time of
recombination $t_{rec}$ to today. The dotted lines show two photon paths
propabating from the last scattering surface to the observer at {\bf O}.
\vskip .5cm
{\bf Figure 2 :} The predicted angular autocorrelation function $C(\alpha)$,
in units of $\mu_6^2$ ($\mu K)^2$.
\vskip .5cm
{\bf Figure 3:} Probability distribution $P({{\delta T}\over T})$ as predicted
by cosmic strings in
the context of the analytic model (3a). The difference from the corresponding
Gaussian distribution
is shown in Fig. 3b. The variable ${{\delta T}\over T}$ is proportional but
not equal to the predicted temperature fluctuations.
\vskip .5cm
{\bf Figure 4:} The analytically obtained probability distribution (tilted
squares) for the {\it
gradient} of ${{\delta T}\over T}$ for an experiment with resolution 18'',
superimposed with the
corresponding Gaussian distribution (squares). The analytical result was also
verified by a
Monte-Carlo simulation (crosses).
\vskip .5cm
{\bf Figure 5 :}  The  predicted probability distribution of the kurtosis for
the VLA
experiment (short dashed line) superimposed with the corresponding distribution
for Gaussian
underlying perturbations (continous line). A similar superposition is also
shown
for
80'' resolution in the same experiment (dotted vs long dashed lines).

\end{flushleft}
\vspace{0.8cm}
\centerline{\bf References}
\vspace{0.5cm}
\begin{flushleft}
Albrecht A. \& Turok N. 1989, Phys. Rev. {\bf D40}, 973.\\
Allen B. \& Shellard E. P. S. 1990, Phys.Rev.Lett. {\bf 64}, 119.\\
Bennett D. \& Bouchet F. 1988, Phys.Rev.Lett. {\bf 60}, 257.\\
Bennett D., Stebbins A. \& Bouchet F. 1992, Ap.J.(Lett.) {\bf 399},
              L5.\\
Brandenberger R. 1992, {\it `Topological Defect Models of Structure Formation
\\
....  After the COBE Discovery of CMB Anisotropies'}, publ. in proc. of the\\
....  International School of Astrophysics "D.Chalonge", 6-13 Sept.1992,\\
....  Erice, Italy, eds. N.Sanchez and A. Zichichi (World Scientific,\\
....  Singapore, 1993).\\
Coulson D., Ferreira P., Graham P. \& Turok N. 1993, `$\Pi$ in the Sky?\\
.... Microwave Anisotropies from Cosmic Defects', PUP-TH-93/1429 (1993).\\
Fomalont E. et al. 1993, Ap.J. {\bf 404}, 8.\\
Graham P., Turok N., Lubin P., Schuster J. 1993, PUP-TH-1408 (1993).\\
Gott J. et al. 1990, Ap.J. {\bf 352}, 1.\\
Hara T. \& Miyoshi S. 1993, Ap. J. {\bf 405}, 419.\\
Kaiser N. \& Stebbins A. 1984, Nature {\bf 310}, 391.\\
Kibble T. W. B. 1976, J.Phys. {\bf A9}, 1387.\\
Luo X. 1993, `Statistical Tests for the Gaussian Nature of Primordial\\
.... Fluctuations Through CBR Experiments', Fermilab preprint (1993).\\
Magueijo J. 1992, Phys.Rev. {\bf D46}, 1368.\\
Moessner R., Perivolaropoulos P. \& Brandenberger R. 1994, Ap. J. {\bf 425},\\
.... April 10, 1994 (to appear).\\
Perivolaropoulos L. 1993a, Phys.Lett. {\bf B298}, 305.\\
Perivolaropoulos L. 1993b, Phys. Rev. {\bf D48} 1530. \\
Perivolaropoulos L. 1993c, {\it `The Fourier Space Space Statistics of \\
.... Seedlike Cosmological Perturbations'},CfA preprint No. 3591, M.N.R.A.S.\\
.... in press (1993).\\
Perivolaropoulos L. \& Vachaspati T. 1993, `Peculiar Velocities and \\
.... Microwave Background Anisotropies from Cosmic Strings', submitted to \\
.... Ap. J. (1993).\\
Perivolaropoulos L., Brandenberger R. \& Stebbins A. 1990, Phys.Rev. {\bf D41},
\\
.... 1764.\\
Smoot G.{\it et. al.}, {\it Ap. J. Lett.} {\bf 396}, L1 (1992).\\
Stebbins A. et al. 1987. Ap. J. {\bf 322}, 1.\\
Stebbins A. 1988, Ap.J. {\bf 327}, 584.\\
Traschen J., Turok N. \& Brandenberger R. 1986, Phys.Rev. {\bf D34},
              919.\\
Turok N. \& Brandenberger R. 1986, Phys. Rev. {\bf D33} 2175.\\
Vachaspati T. 1992, Phys.Lett. {\bf B282}, 305.\\
Veeraraghavan S. \& Stebbins A. 1990, Ap.J. {\bf 365}, 37.\\
Vilenkin A. 1985, Phys.Rep. {\bf 121}, 263.\\
\end{flushleft}
\end{document}